\documentclass[prb,noshowpacs,superscriptaddress,twocolumn]{revtex4-1}

\usepackage{amsmath}
\usepackage{graphicx}
\usepackage{bm}
\usepackage[colorlinks]{hyperref}

\newcommand{\spl}{$\sigma_+$ }
\newcommand{\sm}{$\sigma_-$ }

\begin{document}

\title{Topological turbulence in spin-orbit-coupled driven-dissipative quantum fluids of light generates high angular momentum states}

\author{S.~V.~Koniakhin}
\email{kon@mail.ioffe.ru}
\affiliation{Institut Pascal, PHOTON-N2, Universit\'e Clermont Auvergne, CNRS, SIGMA Clermont, F-63000 Clermont-Ferrand, France}
\affiliation{St. Petersburg Academic University - Nanotechnology Research and Education Centre of the Russian Academy of Sciences, 194021 St. Petersburg, Russia}

\author{G.~Malpuech}
\affiliation{Institut Pascal, PHOTON-N2, Universit\'e Clermont Auvergne, CNRS, SIGMA Clermont, F-63000 Clermont-Ferrand, France}

\author{D.~D.~Solnyshkov}
\affiliation{Institut Pascal, PHOTON-N2, Universit\'e Clermont Auvergne, CNRS, SIGMA Clermont, F-63000 Clermont-Ferrand, France}
\affiliation{Institut Universitaire de France (IUF), 75231 Paris, France}

\author{A.V.~Nalitov}
\affiliation{Institut Pascal, PHOTON-N2, Universit\'e Clermont Auvergne, CNRS, SIGMA Clermont, F-63000 Clermont-Ferrand, France}
\affiliation{Faculty of Science and Engineering, University of Wolverhampton, Wulfruna St, Wolverhampton WV1 1LY, United Kingdom}

\date{\today}

\begin{abstract}
We demonstrate the formation of a high angular momentum turbulent state in an exciton-polariton quantum fluid with TE-TM Spin-Orbit Coupling (SOC). The transfer of particles from quasi-resonantly cw pumped \spl component to \sm component is accompanied with the generation of a turbulent gas of quantum vortices by inhomogeneities. We show that this system is unstable with respect to the formation of bogolons at a finite wave vector, controlled by the laser detuning. In a finite-size cavity, the domains with this wave vector form a ring-like structure along the border of a cavity, with a gas of mostly same-sign vortices in the center. The total angular momentum  is imposed by the sign of TE-TM SOC, the wave vector of instability, and the cavity size. This effect can be detected experimentally via local dispersion measurements or by interference. The proposed configuration thus allows simultaneous experimental studies of quantum turbulence and high-angular momentum states in continuously-pumped exciton-polariton condensates.
\end{abstract}

\maketitle

\section{Introduction}

Quantum turbulence (QT) is a topic of high importance for fundamental theoretical \cite{bradley2012energy,simula2014emergence,valani2018einstein} and experimental \cite{roche2007vortex,Johnstone2019,Gauthier2019} condensed matter physics. It plays an important role in such closely related phenomena as superfluity \cite{bogoliubov1947theory} and superconductivity \cite{bardeen1957theory}, whose studies contributed drastically to the methods of modern theoretical condensed matter physics. Similarly to the turbulence in classical fluids, QT is a complex semi-stochastic motion of matter accompanied with the formation of vortices. The key specificity of QT consists in quantization of vortices due to the restrictions on the phase of the order parameter (the quantum fluid wave function). The QT in multi-component  condensates is of a particular interest, due to the interplay of the spinor degrees of freedom and the emergence of new topological defects, such as  half-vortices  \cite{VolovikReview,Rubo2007,lagoudakis2009observation,SST2010,tsubota2017numerical}. 

A particularly interesting example of a spinor quantum fluid is a macroscopically populated state of exciton-polaritons, light-matter quasiparticles emerging in optical microcavities in the strong coupling regime \cite{Microcavities2017}. Such quantum fluids can either be created by Bose-Einstein condensation in an equilibrium configuration \cite{kasprzak2006bose,Kasprzak2008}, or in a highly-nonequilibrium conditions by direct quasi-resonant pumping \cite{lagoudakis2008quantized,amo2011polariton}.
The wave function of the polariton fluid can be controlled directly by resonant optical excitation through the photonic component, allowing the generation of topological defects in polariton quantum fluids: solitons \cite{amo2011polariton,claude2020taming,lerario2020parallel} and half-solitons \cite{hivet2012half}, quantum vortices \cite{lagoudakis2008quantized,dominici2015vortex,boulier2015vortex} and half-vortices \cite{lagoudakis2009observation,dominici2015vortex}, vortex chains\cite{hivet2014interaction,soln2019vortexstreet}, and even analog black holes \cite{nguyen2015acoustic,solnyshkov2019quantum}. At the same time, repulsive spin-anisotropic polariton interactions \cite{Renucci2005,Vladimirova2010} and bosonic stimulated scattering \cite{Savvidis2000} allow indirect control over polariton condensates through non-resonant generation of reservoir excitons, providing both gain and effective potential \cite{Wouters2007a,sanvitto2011all,Dreismann2014b}. 

This all-optical control over the order parameter of polaritons is especially important in the context of turbulence generation \cite{soln20202turb}.
Rotating potentials can be used for stirring polariton quantum fluids as it is done with liquid Helium \cite{Andronikashvili1946,rusaouen2017detection} and atomic condensates  \cite{Madison2000,White2014}.  Even with a constant potential, its interplay with the gain-loss competition is sufficient for the generation of stable high angular momentum condensates \cite{Dreismann2014b,Sun2018,Nalitov2019}. High angular momentum polariton condensates can also be generated by resonant driving \cite{bramati4lasers,boulier2015vortex,gavrilov}. Polariton condensates are naturally emitting coherent light, and such high angular momentum beams have various applications \cite{zhao2020measuring,ruffato2019multiplication} including high-resolution microscopy \cite{ritsch2017orbital}, quantum information \cite{mirhosseini2015high}, and micromanipulation \cite{shen2018polygonal}.
    
Spinor polariton condensates are subject to TE-TM SOC which shows up as the energy splitting of transverse-electric and transverse-magnetic optical modes of the cavity and results in the optical spin Hall effect \cite{kavokin2005optical,leyder2007observation,liew2007,Kammann2012,nalitov2015spin,sheffield2020}. The TE-TM effective field couples the two pseudospin components of spinor polariton condensates, affects the topological defects \cite{Rubo2007,Flayac2010,hivet2012half}, and leads to the non-trivial topology of the polariton bands \cite{gianfrate2020measurement} in the presence of an external magnetic field, either real or effective \cite{nalitov2015polariton,Bleu2016,Bleu2017,Klembt2018,Sigurdsson2019}. Laser emission with non-zero angular momentum arising thanks to the TE-TM SOC has also been demonstrated recently \cite{zambon2019optically}.

In this work, we demonstrate that the interplay of the topological excitations of a spinor quantum fluid (half-vortices) with the weak excitations (bogolons) can lead to the development of the high angular momentum states from topological quantum turbulence via instabilities.  We consider a circularly-polarized resonant pumping of a large pillar microcavity, above the bistability threshold. The self-induced Zeeman splitting splits the bands at $k=0$. At higher $k$, the two circular spins are coupled by the TE-TM SOC. Both bands show a non-zero Berry curvature of opposite signs. First, the spin conversion takes place because of the spatial inhomogeneity of the pump creating non-zero wave vectors sensitive to TE-TM and exhibiting the optical spin Hall effect. This spin conversion is known to generate vortices \cite{liew2007} because of the non-trivial band topology \cite{Chang2008}, but the numbers of generated vortices and anti-vortices remain almost equal. Quantum turbulence is thus achieved in an energy band with a non-zero Berry curvature. Second, a bogolon instability leads to the amplification of a parametric process, in which polaritons from the pumped $k=0$ state in the \spl component are scattered towards a $k$ state of the \spl-like branch and toward a $-k$ state of the \sm-like branch. This process is energetically resonant and possible thanks to the TE-TM SOC. The isotropic instability amplifies the small symmetry breaking provided by the initial vortex-antivortex imbalance, leading to the generation of a macroscopic rotating current along the system boundary, with a turbulent gas of mostly same-sign vortices accumulated in the center. We show that the formation of such states can be detected experimentally via interference or dispersion measurements.

\begin{figure}[tbp]
\begin{center}
    \includegraphics[width=0.72\linewidth]{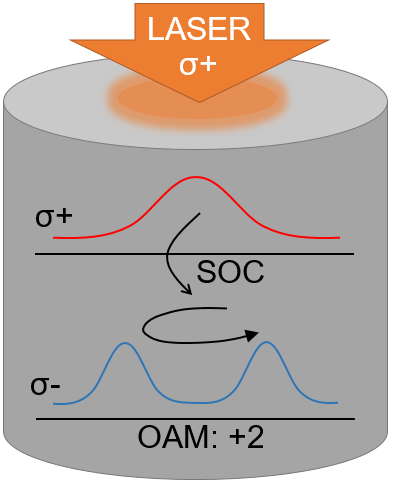}
\end{center}
\caption{Scheme of angular momentum transfer and vortex creation in the spinor polariton quantum fluid with spin orbit coupling stemming from TE-TM splitting.}
\label{fig_scheme}
\end{figure}

\begin{figure}[tbp]
\includegraphics[width=0.9\linewidth]{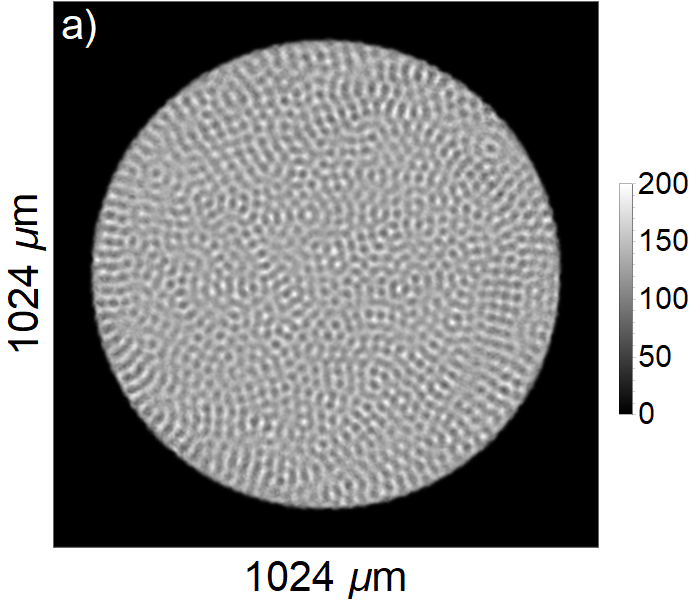}\\
\includegraphics[width=0.9\linewidth]{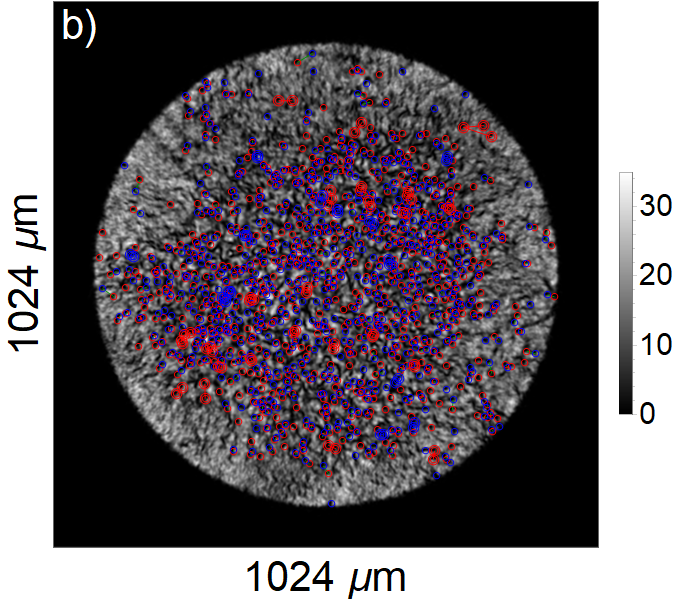}
\caption{Panel a) Density in $\mu$m$^{-2}$ of $\sigma^+$ polarization component. Panel b) Density of $\sigma^-$ polarization component with the detected vortices and their clusters.}
\label{fig_12comp_dens}
\end{figure}

\begin{figure}[tbp]
\includegraphics[width=0.9\linewidth]{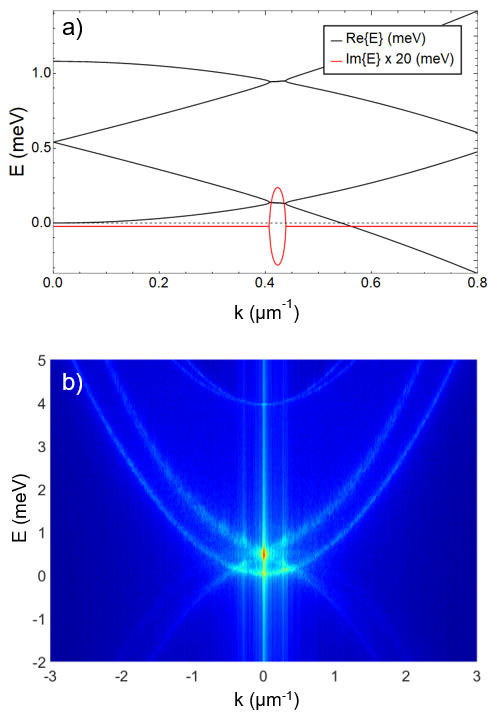}
\caption{Panel a) Dispersion of DDGPE linearized Bogoliubov-like excitations obtained via BdG equations for the system of $\sigma^+$ polarization component in sufficiently non-linear regime and $\sigma^-$ component with lower density. Panel b) The dispersion from all space obtained from the numerical simulations. The sum of signals from $\sigma^+$ and $\sigma^-$ is given.}
\label{fig_BdG_and_disp_matlab}
\end{figure}

\begin{figure*}[tbp]
\includegraphics[width=0.99\linewidth]{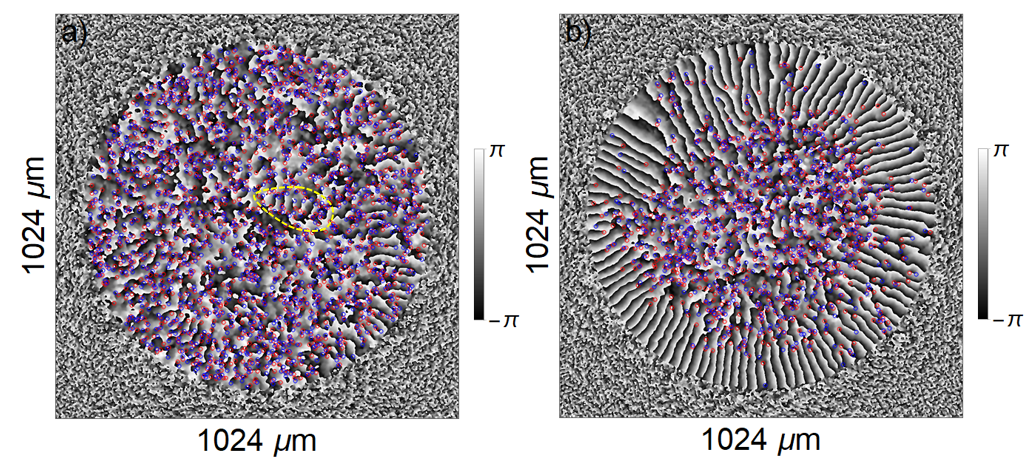}\\
\includegraphics[width=0.49\linewidth]{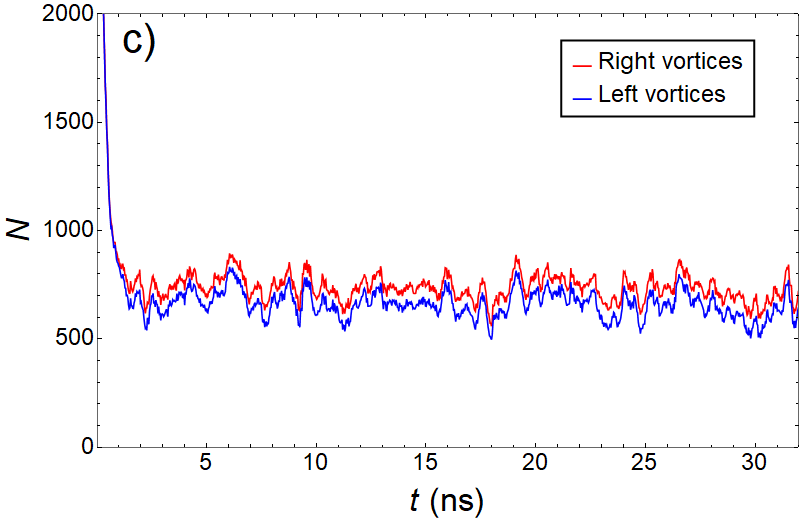}\includegraphics[width=0.49\linewidth]{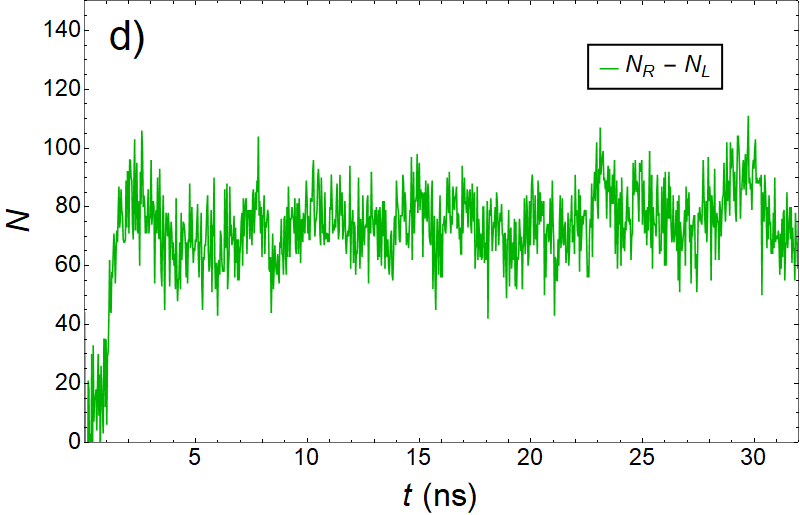}
\caption{Panels a,b) are for the phase of \sm component at 1 ns and 29 ns, respectively. Yellow dashed curve circles the domain of wave vector $k_0$ spontaneously appeared due to the instability. Panels c) and d) show the time dependencies of right/left vortices numbers and of their difference, respectively.}
\label{fig_phase_and_time_depend}
\end{figure*}

\begin{figure*}[tbp]
\includegraphics[width=0.99\linewidth]{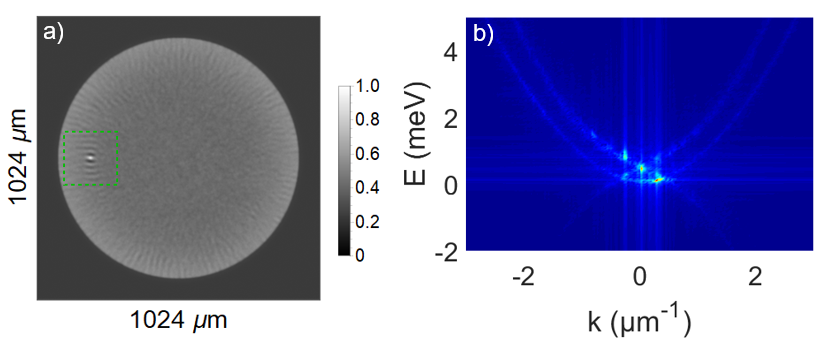}
\caption{Panel a) Time-averaged over 25 ns interference pattern $|\psi(\mathbf{R}_0)+\psi(\mathbf{r})|^2$ of some fixed point $\mathbf{R}_0$ with all other points $\bf r$. The point $\mathbf{R}_0$ is visible as bright spot. Fringes show the presence of a flow with $k \approx 0.3~\mu$m$^{-1}$, which matches with instability regions of linearized excitations. Panel b) Dispersion from the region marked by dashed square in panel a). The dominating direction is well visible.}
\label{fig_exper}
\end{figure*}


\section{Spin conversion and vortex generation by the TE-TM SOC}


The non-equilibrium exciton-polariton condensates combine spinor and non-Hermitian physics. They can be described by the Driven-Dissipative Gross-Pitaevskii Equation (DDGPE), valid for polaritons at sufficiently small wave vectors:
\begin{multline}
    i\hbar\frac{\partial \psi_{\pm} }{\partial t}  = \left[-\frac{\hbar^2\nabla
^2}{2m} -i \Gamma + g\left| \psi_{\pm}\right|^2 + V({\bf r}) \right]\psi_{\pm} \\
+ P_{\pm} e^{-i \omega_0 t} + \beta (\partial_x \pm i \partial_y)^2 \psi_{\mp}.
\label{eq_GPEph}
\end{multline}
Here, $\psi_{\pm}$ are the two pseudospin components of the condensate wavefunction, corresponding to $\sigma_+$ and $\sigma_-$ circular polarizations of the emitted photons, $m$ is the polariton effective mass, $2\Gamma$ is the inverse polariton lifetime, $g$ is the strength of spin-anisotropic polariton interaction (we neglect the interaction between the polaritons of opposite spins), and $\beta$ is the strength of the TE-TM effective field, coupling the two spinor components.

In our calculations, the pumping of the \spl-component $P_+$ is spatially inhomegeneous, with an average value $P_0$ (a Gaussian-filtered white noise with a rms amplitde of $0.28P_0$, a correlation length 3 $\mu$m and biased by $P_0$) and the pumping of the other component is absent $P_- = 0$. The magnitude of $P_0$ is slightly higher than the  bistability threshold. The rotationally-symmetric confining potential defined via the Heaviside function as $V({\bf r}) = 4~{\rm meV}  \cdot H(|{\bf r}|-L)$ is used, where $L=400~\mu{\rm m}$. The other experimentally relevant parameters read $\Gamma = 1/2\tau_{\rm pol}$ with $\tau_{\rm pol} = 300$~ps, $\hbar \omega_0 = 0.54$~meV, $m = 5\cdot 10^{-5}m_0$ with $m_0$ the free electron mass, $g=5~\mu$eV$\mu$m$^{-2}$, and $\beta = \hbar^2/4\cdot(m_l^{-1}-m_t^{-1})$ with $m_l=m_0=0.75m_t$. In numerical simulations, we used the third order Adams–Bashforth scheme in time and calculated the Laplace operator for kinetic energy via the Fourier transform using the Graphic Processor Unit acceleration in Matlab package.

We begin with the description of the vortex generation during the spin conversion by the TE-TM SOC. The inhomogeneous pumping creates an inhomogeneous density profile in the \spl component. The same profile can be achieved by disorder. A single maximum of the condensate density can be considered as a source of a divergent polariton flow (see Fig.~\ref{fig_scheme}). The configuration is thus quite similar to that of the optical spin Hall effect \cite{kavokin2005optical}, known to generate vortices \cite{liew2007} by spin conversion. Divergent effective fields (generated here by the wave vector-dependent TE-TM SOC) were used to create and destroy (unwind) vortices in spinor condensates a long time ago \cite{Matthews1999,LeggettBook}. Later, vortex generation by spin-orbit coupling has been linked with the non-trivial band topology characterized by the non-zero Berry curvature \cite{Chang2008}. 
In our particular case, the divergent currents are amplified by the spin-anisotropic polariton-polariton interactions, which create a potential barrier, repelling and accelerating \spl polaritons. Each density maximum generates a pair of vortices of a sign determined by the winding of the TE-TM SOC (two vortices are generated because of the winding number 2). However, each density minimum, acting as a point of convergent currents, also generates a pair of vortices of the opposite sign. In total, the number of vortices and antivortices is almost the same, but the symmetry is broken by the fact that the system is confined in a finite-size pillar (even if it is very large).

We now present the results of numerical simulations confirming the above analysis. Figure~\ref{fig_12comp_dens} shows the densities in \spl and \sm components after 32~ns evolution. As expected, the density of the pumped \spl component is inhomogeneous, but it does not exhibit any topological defects, whereas the second component \sm is much more inhomogeneous. While the grayscale of the figure shows only the particle density, we have also used the phase of the calculated wave function in order to detect the vortices (red and blue circles) according to the automatic procedure described in \cite{soln20202turb}. A lot of vortices of both signs are visible in this figure. They form a random pattern. The system does not converge to a stationary solution, and the wavefunction changes with time. The associated vortex motion represents a serious problem for their experimental detection \cite{boulier2015vortex,soln20202turb}. On the other hand, it confirms that the phase of the \sm component is not fixed by the \spl pump, and the vortices are free. This gas of freely moving vortices in a single spin component represents a particular implementation of the quantum turbulence in a topologically non-trivial band.

\section{Bogolon instability}

We now turn to the study of the interplay of the topological excitations (vortices, or rather half-vortices, since they are present in a single spin component) with weak collective excitations of the spinor condensate (bogolons). We analyze the stability of the system with respect to weak perturbations with a certain wave vector $\bf k$ using the Bogoliubov-de Gennes equations~\eqref{eq_bdgSystem} with the assumption that the density of the pumped component \spl is much higher than the density of the other component \sm  ($g|\psi_-|^2 \ll \hbar \omega_0\approx g|\psi_+|^2$). The presence of the vortices in the \sm component and the spatial inhomogeneity of the \spl component are neglected at this scale. Similar problems for spinor condensates were considered previously\cite{rubo2006suppression,solnyshkov2008disp,kovalev2017paramagnetic}, but never under quasi-resonant circular polarized pumping in presence of the TE-TM splitting.  These equations represent an eigenvalue problem with respect to the bogolon energy $\hbar \omega$. The following notation is used: $\hat{T}$ denotes the kinetic energy operator, $\psi_{+A}, \psi_{+B}, \psi_{-A}, \psi_{-B}$ are the small perturbations of the wave function, $\phi$ is the polar angle of the wave vector $\bf k$. 

\begin{widetext}
\begin{equation}
\left( {\begin{array}{*{40}{c}}
{ \hat{T} + 2g {{\left| {{\psi _+}} \right|}^2} - \hbar {\omega_0} }&{ - g \psi _+^2} & \beta k^2 e^{2i\phi} & 0\\
{g \psi _+^{*2}}&{ - \hat{T} - 2g {{\left| {{\psi _+}} \right|}^2}  + \hbar {\omega_0} }& 0 & -\beta k^2 e^{-2i\phi}\\
\beta k^2 e^{-2i\phi} & 0 & \hat{T} - \hbar {\omega_0} & 0\\
0 &-\beta k^2 e^{2i\phi} &  0 & - \hat{T} + \hbar \omega_0
\end{array}} \right)
\left( {\begin{array}{*{40}{c}}
\psi_{+A}\\
\psi_{+B}\\
\psi_{-A}\\
\psi_{-B}
\end{array}} \right) = (\hbar \omega + i \Gamma) \left( {\begin{array}{*{40}{c}}
\psi_{+A}\\
\psi_{+B}\\
\psi_{-A}\\
\psi_{-B}
\end{array}} \right).
\label{eq_bdgSystem}
\end{equation}
\end{widetext}

Due to the rotational symmetry of the system in the circular polarization basis, the eigenvalues of Eq.~\eqref{eq_bdgSystem} do not depend on $\phi$. The analytical solution is given by Eq.~\eqref{sol}.
\begin{widetext}
\begin{eqnarray}\label{sol}
    E&=&-i\Gamma\pm\frac{\sqrt{2}}{2}\left[\left(\hbar\omega_0\right)^2+2\left(\left(\frac{\hbar^2 k^2}{2m}\right)^2+\left(\beta k^2\right)^2\right)\right.\\
    &\pm&\left.\sqrt{\left(\hbar\omega_0\right)^4-8\left(\frac{\hbar^2 k^2}{2m}\right)\left(\hbar\omega_0\right)^3+\left(16\left(\frac{\hbar^2 k^2}{2m}\right)^2\left(\hbar\omega_0\right)^2-4\left(\beta k^2\right)^2\right)+16\left(\frac{\hbar^2 k^2}{2m}\right)^2\left(\beta k^2\right)^2}\right]^{1/2}\nonumber
\end{eqnarray}
\end{widetext}

Fig.~\ref{fig_BdG_and_disp_matlab} shows the spectrum of Bogoliubov excitations obtained from Eqs.~\eqref{eq_bdgSystem} and the numerically calculated spectrum, obtained by Fourier-transforming the solution of Eq.~\eqref{eq_GPEph}. The signatures of the bogolon instability are clearly visible in both panels. The analytical solution presents a region with a flat real part and a positive imaginary part of the energy of the weak excitations, which appears in the region where the branches of the dispersion of the two spin components cross each other. This behavior is typical for Bogoliubov-de Gennes equations because of their non-Hermitian form. Contrary to the previous studies \cite{solnyshkov2008disp}, where such crossings did not give rise to instabilities, here the bands are not purely circular-polarized because of the TE-TM SOC, which is precisely what allows  the resonant parametric processes at these crossing points to take place by providing the corresponding eigenvectors a finite overlap. The predictions of the analytical calculations based on the linearized model are completely confirmed by the full numerical simulations, where the bright spots visible at the crossing points of the dispersions of the two spin components correspond to the same instability regions.

As a consequence, the bogolon states with these wave vectors exhibit exponential growth, which acts as an additional source of the spin conversion, bringing particles with the associated wave vectors into the second spin component \sm. The highest growth rate wave vector of instability $k_0$ can be estimated from the condition $\hbar \omega_0 \approx \hbar k_0 c_+$, where $c_+ = \sqrt{\hbar \omega_0/m}$ is the sound velocity in the \spl component. Here, one neglects the contribution from the parabolic polariton dispersion in the \sm component with respect to the linear bogolon dispersion in the \spl component. This rate is relatively slow. Moreover, in a perfect system the bogolon dispersion is cylindrically symmetric, which means that there is a mode competition for this instability, which reduces the development of a particular mode. In what follows, we will see that the symmetry breaking provided by the topological quantum turbulence allow one particular mode to win this competition and to develop a macroscopic current.


\section{Formation of the high angular momentum state}

We now consider the interplay of the topological defects and the bogolon instability. The random motion of the vortices in the \sm component allows the formation of vortex-free domains, and the bogolon instability favors a particular absolute value of the wave vector within any of them. Fig.~\ref{fig_phase_and_time_depend}a) shows a snapshot of the condensate phase at this early stage with one of such domains highlighted. The automatically detected vortices appear as red and blue circles. Because of the non-zero wave vector, these domains must move, and because of the irrotationality of the quantum fluid, their motion is controlled by the algebraic sum of the winding numbers of all vortices. When such a domain forms close to the boundary of the system, it expels the vortices on its external side towards this boundary, where they can disappear without violating the topological constraints. In this process, vortices of one sign disappear more often than the vortices of the other sign, because their positions are correlated with the propagation direction of the domain. In the example shown in Fig.~\ref{fig_phase_and_time_depend}(a), there are more "red" vortices in the central part of the system and more "blue" vortices between the vortex-free domain and the boundary, which is why the number of the "blue" vortices decreases faster than the number of the red ones. The reduction of the vortex density allows the domain to grow. The finite result of this growth is shown in Fig.~\ref{fig_phase_and_time_depend}(b), where a large domain with a clockwise circulating current appears along the whole boundary of the system. This can be seen as a result of a mode competition, with the clockwise direction being favored over the anti-clockwise by the small initial difference in the number of vortices (itself being determined by the winding of the TE-TM and the real-space topology of the cavity). We note that the flow inside the vortex-free domain is supersonic, because the interactions in the \sm component are weak: the velocity attributed to $k_0$ ($\approx 10^8$~cm/s) is higher than the sound velocity in the high-density regions ($\approx 5\cdot 10^7$~cm/s).
 
 The evolution of the vortex populations is shown in Fig. \ref{fig_phase_and_time_depend}c), which presents the numbers of vortices and anti-vortices versus time, and Fig. \ref{fig_phase_and_time_depend}d), which shows the difference of the two. The duration of the transitional regime is 1.5-2~ns and then the stationary regime starts. During this stage, the total number of vortices decreases from several thousands to $N_+ + N_- \approx 1500$. The difference in the number of right and left vortices on the contrary increases and reaches $\Delta N = N_+ - N_- \approx 75$.  Interestingly, the difference of the opposite vortex numbers can be estimated as $\Delta N \approx 2\pi L k_0$.
It follows from the fact that the annular flow with a local wave vector $k_0$ 
around a circle with the radius $L$ corresponds to an angular momentum $M\approx 2\pi L k_0$ (in units of $\hbar$, from the classical definition of the angular momentum), but in a quantum fluid, which is irrotational, all this angular momentum has to be provided by vortices, which determines their net winding number $\Delta N$. The total amount of vortices may be estimated as $N\approx L^2/\xi^2$, where $\xi \approx 3.6~\mu$m is the healing length obtained for the high-density regions in the \sm component ($\approx 20~\mu$m$^{-2}$). These results confirm our interpretation of the process.

In the stationary regime, the first component density and the phase (with respect to the laser) slightly fluctuate around their mean values, while in the second component one still observes very rich behavior. The intervortex distance is comparable with the healing length, the vortices are clearly distinguishable by the phase jumps, however, from the density distribution this vortex gas resembles a mixture of vortices and solitons/deep bogolons (for vortex-antivortex pairs). We note that while in general, one can expect the condensation of vortices to occurs at particular temperatures \cite{valani2018einstein}, in the present configuration no signatures of this effect were detected.

The whole simulation videos with 32 ns duration are available online (in supplementary materials):
\begin{itemize}
    \item Phase of the \sm component \href{https://youtu.be/5S8k9spt6ck}{https://youtu.be/5S8k9spt6ck}
    \item Density of the \sm component and its Fourier image \href{https://youtu.be/mmfRtps8cH8}{https://youtu.be/mmfRtps8cH8}
\end{itemize}


\section{Experimental signatures}


The turbulent regime is particularly challenging for studies in polariton systems, requiring single-shot experiments \cite{soln20202turb}. Indeed, the stochastic nature of the turbulence precludes the usual pulsed experiments based on the stroboscopic principle of the streak cameras. The vortex dynamics can be measured only if it is repeated exactly the same way with each pulse \cite{dominici2015vortex}, otherwise the vortex trajectories and the phase patterns are smeared out by averaging.

The present configuration presents a particular interest, because, in spite of the turbulent vortex gas present in the center of the pillar, it allows macroscopic measurements of the net angular momentum in the cw regime, without any need for pulsed or single-shot experiments. Two approaches can be used. One of the options is measuring interference patterns produced by superposition of the condensate emission with emission of its single point near the cavity border. This scheme is similar to the off-axis interference measurements performed in Refs. \cite{bramati4lasers,boulier2015vortex}. The period of interference fringes at the cavity circumference yields the wave vector $k_0$, see Fig. \ref{fig_exper}a), obtained by calculating the time-averaged self-interference $|\psi(\mathbf{R}_0,t)+\psi(\mathbf{r},t)|^2$ with the wave function calculated numerically with Eq.~\eqref{eq_GPEph}. We see that the interference pattern in the vicinity of the reference point $\mathbf{R}_0$ is not smeared out in spite of the time averaging. This occurs because the vortices are concentrated in the central part of the system, and the reference point is situated in the vortex-free domain.

However, this scheme does not allow to determine the direction of the flow. Alternatively, both the absolute value and the sign of the net condensate topological charge can be deduced from the local angle-resolved measurements of polariton emission spectrum, as shown in Fig. \ref{fig_exper}b), obtained by Fourier-transforming the wave function $\psi(\mathbf{r},t)$. Extracting the wave vector $k_0$ of the domain from the position of the brightest emission points in the reciprocal space allows calculating the total angular momentum as a product of $L$ and $k_0$, and thus deducing the net vorticity in the central region.

\section{Conclisions}

The key role in the generation of the high angular momentum states is played by the OPO-like process of resonant transfer of particles between the \spl and \sm components, based on the instability at a certain wave vector, obtained from the Bogoliubov-de Gennes equations. The resonant nature of the particle transfer leads to the fact that this effect can be observable at sufficiently low values of disorder strength in \spl component. Experimentally, the desired disorder, sufficient to trigger the effect, is usually naturally present in the pumping laser intensity and the cavity detuning. At the same time, the effect is observable only at the \spl component pumping laser amplitude near in the right part of the bistability hysteresis loop or slightly above the loop. At lower intensities, the particle density is too small, while at higher densities, the \spl component is too homogeneous due to the phase pinning effect.

The final configuration of the \sm component corresponds to a gas of free vortices, dominated by the vortices of a given sign defined by the TE-TM SOC. This disequilibrium in the right and left quantum vortices numbers leads to a strong global rotational motion, visible as a circular flow along the border of a circle-shaped cavity. The latter can be detected experimentally either via self-interference experiment or local wave vector-resolved signal in the \sm component. The free motion of the vortices allows referring to this state as to a turbulent state. However, the mutual transfer of particles between \spl and \sm components, as well as the non-conservative nature of the system (driven-dissipative, with quasi-resonant pumping), preclude the observation of a simple -5/3 Kolmogorov law \cite{tsubota2017numerical}. Also, the vortex core is affected by the TE-TM SOC with respect to classical quantum vortex in a single-component quantum fluid\cite{bradley2012energy}. Other deviations from the simple picture of a scalar quantum fluid  were shown in Ref. \cite{voronova2017internal}, where the calculation was using the two component photon-exciton basis. Our results can also be compared with Ref. \cite{gavrilov}, where vortex generation was predicted in two-component polariton quantum fluids, but no SOC coupling was involved and the pumping of the components was equivalent in amplitudes. Finally, in Ref. \cite{yulin2020spinning}, the system was closer to an equilibrium condensate than present one, because of the off-resonant pumping, and the TE-TM SOC was also shown to lead to the formation of rotating structures.

To conclude, we proposed a scheme for the generation of high angular momentum states in spinor polariton quantum fluids from the interplay of the topological quantum turbulence and the bogolon instability.\\

\begin{acknowledgments}

We acknowledge the support of the projects EU Marie Curie ”QUANTOPOL” (846353), ”Quantum Fluids of Light” (ANR-16-CE30-0021), of the ANR Labex GaNEXT (ANR-11-LABX-0014), and of the ANR program ”Investissements d’Avenir” through the IDEX-ISITE initiative 16-IDEX-0001 (CAP 20-25).
\end{acknowledgments}

\bibliography{_BIB}

\end{document}